\input{aipcheck}

\documentclass[
    ,final            
  ]
  {aipproc}

\layoutstyle{6x9}

\newcommand{\msbar}{$\overline{\mbox{MS}}$ }

\begin{document}

\begin{flushright}
CERN-PH-TH/2005-118
\end{flushright}

\title{Impact of large-x 
resummation\\ on parton distribution functions
\footnote{Talk given by G.~Corcella at DIS 2005, 
XIII Workshop on Deep Inelastic Scattering,
April 27--May 1, 2005, Madison, WI, U.\ S.\ A.}
}


\author{G. Corcella}{
  address={CERN, Department of Physics, 
Theory Division, CH-1211 Geneva 23, Switzerland}
}

\author{L. Magnea}{
  address={
Universit\`a di Torino and INFN, Sezione di Torino, Via P.~Giuria 1, I-10125,
Torino, Italy}
}

\begin{abstract}
We investigate the effect of large-$x$ resummation on parton distributions
by performing a fit of Deep Inelastic Scattering data from the NuTeV,
BCDMS and NMC collaborations, using NLO and NLL soft-resummed coefficient
functions. Our results show that soft resummation has a visible impact on
quark densities at large $x$.  Resummed parton fits would
therefore be needed whenever
high precision is required for cross sections evaluated near partonic 
threshold.
\end{abstract}

\maketitle


A precise knowledge of parton distribution functions (PDF's) at large $x$ is
important to achieve the accuracy goals of the LHC and other
high energy accelerators. We present 
a simple fit of Deep Inelastic Scattering (DIS) structure function 
data, and extract NLO and NLL-resummed  quark densities, 
in order to establish qualitatively the effects of soft-gluon 
resummation.

Structure functions $F_i(x,Q^2)$ are given by the convolution of 
coefficient functions and PDF's. Finite-order 
coefficient functions present logarithmic terms that are singular at 
$x = 1$, and originate from soft or collinear gluon radiation. 
These contributions need to be resummed to extend the validity 
of the perturbative prediction.
Large-$x$ resummation for the DIS coefficient function was performed 
in \cite{Sterman:1986aj,Catani:1989ne} in the massless approximation, 
and in \cite{Laenen:1998kp,Corcella:2003ib} with the inclusion of 
quark-mass effects, relevant at small $Q^2$.

Soft resummation is naturally performed in moment space, 
where large-$x$ terms correspond, at ${\cal O}(\alpha_s)$, to 
single ($\alpha_s \ln N$) and double ($\alpha_s \ln^2 N$) logarithms  
of the Mellin variable $N$. In the following, we shall consider values 
of $Q^2$ sufficiently large to neglect quark-mass effects.
Furthermore, we shall implement soft resummation in the next-to-leading
logarithmic (NLL) approximation, which corresponds to keeping terms ${\cal O} 
(\alpha_s^n \ln^{n+1} N )$ (LL) and ${\cal O} (\alpha_s^n \ln^n N)$ 
(NLL) in the Sudakov exponent.

\begin{figure}
\centerline{\resizebox{0.48\textwidth}{!}{\includegraphics{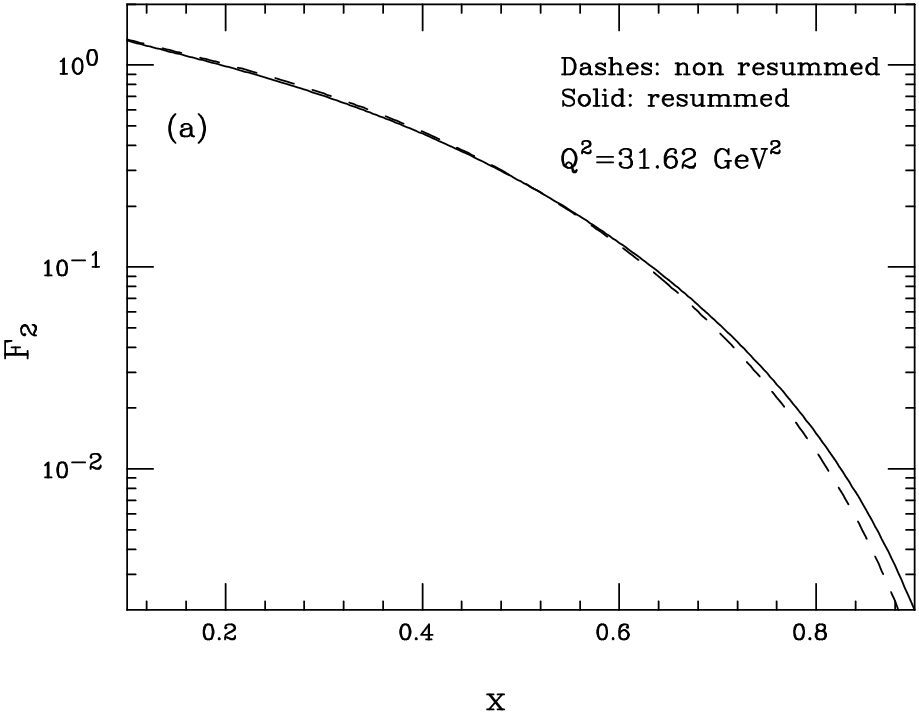}}%
\hfill%
\resizebox{0.48\textwidth}{!}{\includegraphics{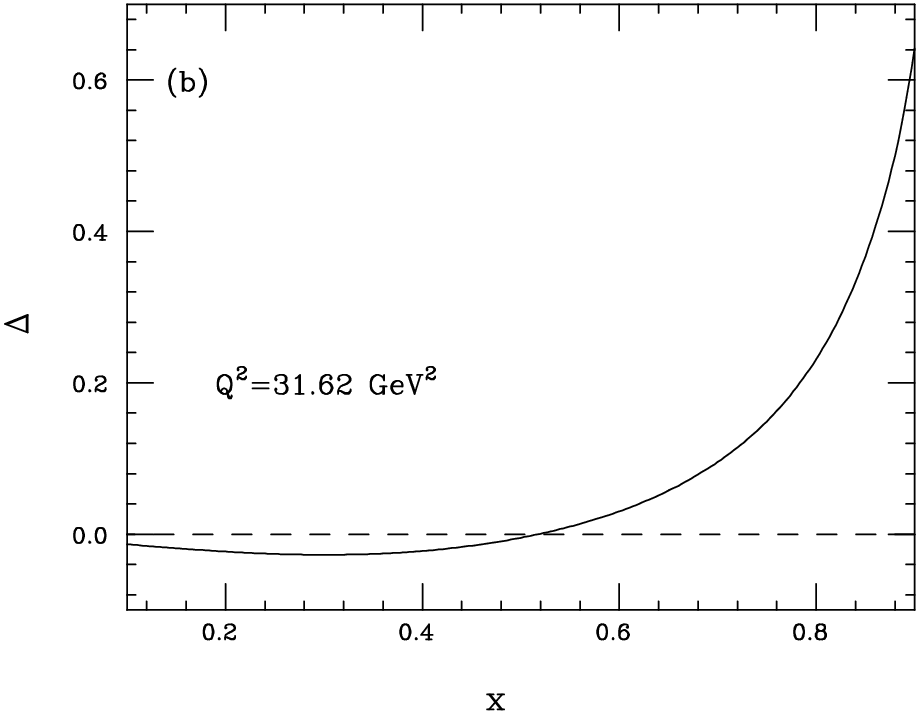}}}
\caption{(a): CC structure function $F_2(x)$ using NLO 
(dashes) and NLL-resummed (solid) coefficient functions, at $Q^2 = 
31.62$~GeV$^2$;
(b): relative difference $\Delta = (F_2^{\mathrm{res}} - 
F_2^{\mathrm{NLO}})/F_2^{\mathrm{NLO}}$}
\label{fdel}
\end{figure}

To gauge the impact of the resummation on the DIS cross section, we can 
evaluate the charged-current (CC) structure function $F_2$ 
convoluting NLO and NLL-resummed \msbar coefficient functions with the NLO 
PDF set CTEQ6M \cite{Pumplin:2002vw}. We consider $Q^2 = 31.62$~GeV$^2$, 
since it is one of the values of $Q^2$ at which the NuTeV collaboration
collected data \cite{Naples:2003ne}. In Fig.~\ref{fdel} we plot $F_2(x)$ with 
and without resummation (Fig. 1a), as well as the normalized difference 
$\Delta = (F_2^{\mathrm{res}} - F_2^{\mathrm{NLO}})/F_2^{\mathrm{NLO}}$ 
(Fig. 1b). 
We note that the effect of the resummation is an enhancement of $F_2$ for
$x > 0.6$. Such an enhancement is compensated by a decrease at smaller $x$: 
the resummation, in fact, does not change the first moment of $F_2$, since 
we include in the Sudakov exponent only terms $\sim \ln^k N$, which vanish 
for $N = 1$.
\begin{figure}[ht!]
\centerline{\resizebox{0.55\textwidth}{!}{\includegraphics{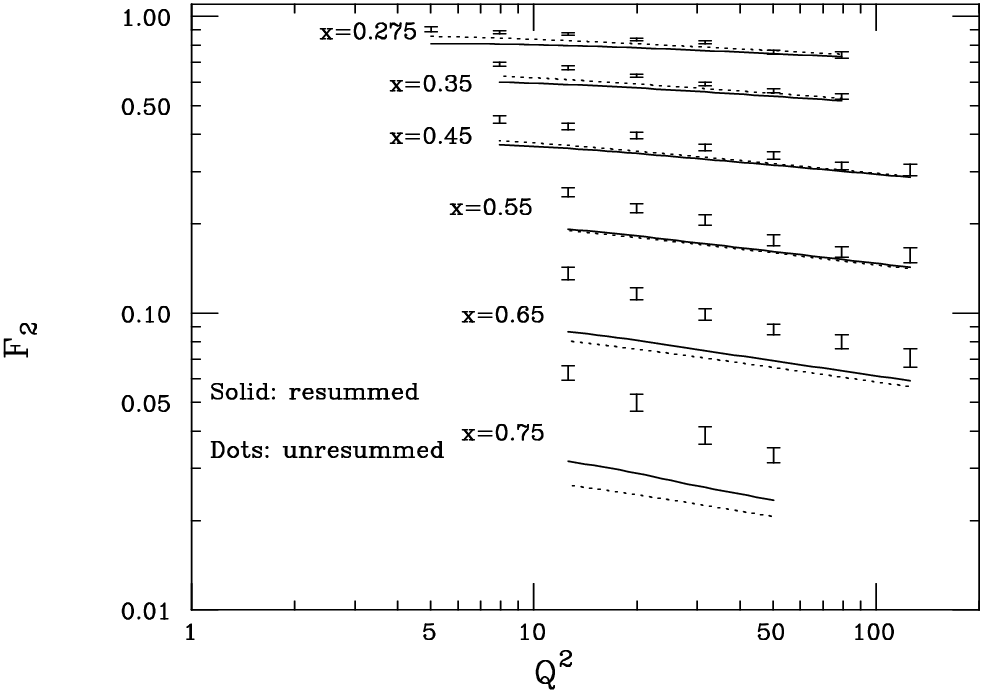}}}
\caption{Comparison of NuTeV data on the CC structure function 
$F_2(x,Q^2)$ with 
a theoretical prediction using CTEQ6M PDF's and NLO (dots)
or NLL-resummed (solid) coefficient functions.}
\label{fignut}
\end{figure}
Our predictions for $F_2$ at different values of $Q^2$ can be compared 
with NuTeV data at large $x$. The results of the comparison are shown in 
Fig.~\ref{fignut}: although the resummation moves the prediction towards 
the data, we are still unable to reproduce the large-$x$ data.
\begin{figure}[ht!]
\centerline{\resizebox{0.48\textwidth}{!}{\includegraphics{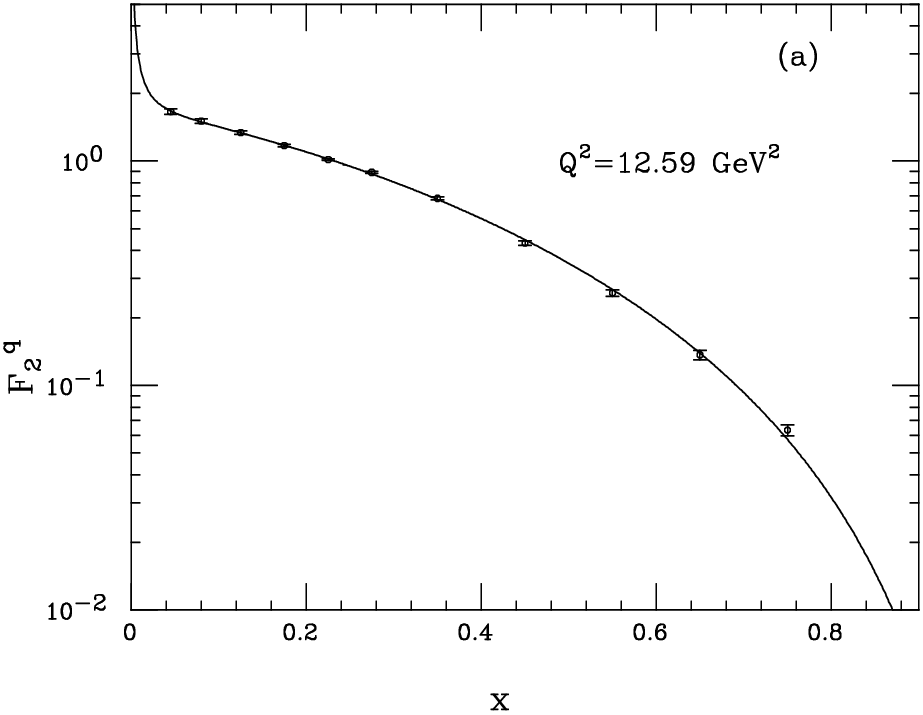}}%
\hfill%
\resizebox{0.48\textwidth}{!}{\includegraphics{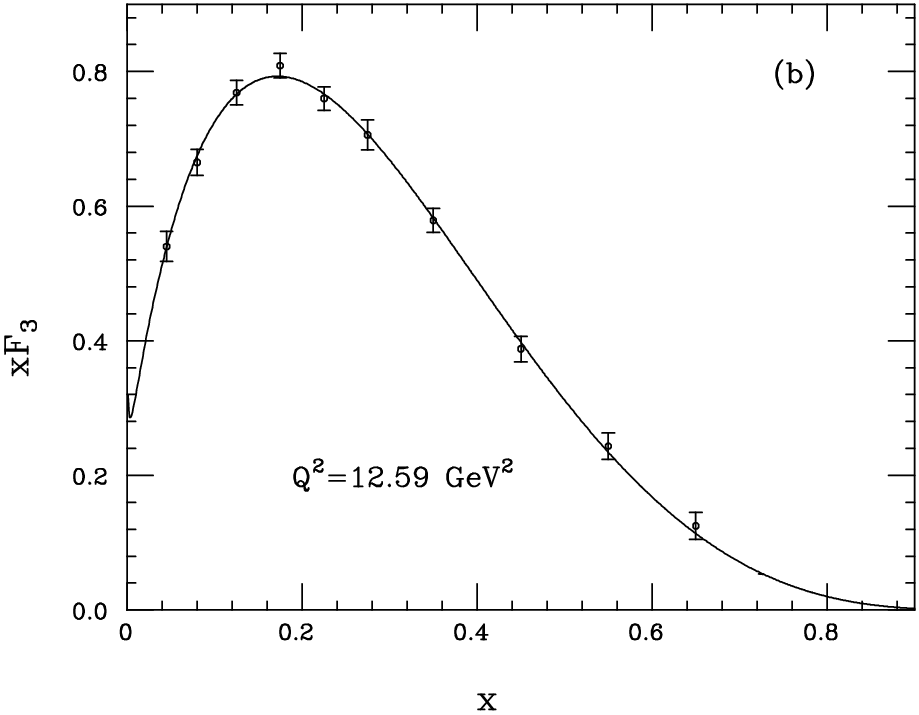}}}
\caption{NuTeV data and best-fit curves at $Q^2=12.59$~GeV$^2$ for
$F_2^q$ (a) and $x F_3$ (b).}
\label{fits}
\end{figure}
Several effects are involved in the mismatch: at very large values of 
$x$, power corrections will certainly play a role. 
Moreover, we have used so far a parton set (CTEQ6M), extracted by a 
global fit which did not account for the NuTeV data. Rather, data from 
the CCFR experiment \cite{Yang:2000ju}, which disagree at large 
$x$ with NuTeV \cite{Naples:2003ne}, were used.
The discrepancy has recently been 
described as understood \cite{tzanov}; however, 
it is not possible to draw any firm conclusion from our comparison.  

We wish to reconsider the CC data in the context of 
an indipendent fit.  We shall 
use NuTeV data on $F_2(x)$ and $x F_3(x)$ at $Q^2 = 31.62$~GeV$^2$  
and 12.59~GeV$^2$, and
extract NLO and NLL-resummed quark distributions from the fit. 
$F_2$ contains 
a gluon-initiated contribution $F_2^g$, which is not 
soft-enhanced and is very small at large $x$: we can therefore safely 
take $F_2^g$ from a global fit, e.g. CTEQ6M, and limit our  
fit to the quark-initiated term $F_2^q$. We choose a
parametrization of the form $F_2^q (x) = F_2 (x) - F_2^g (x) = A 
x^{- \alpha} (1 - x)^\beta (1 + b x)$; $ x F_3(x) = C x^{-\rho} 
( 1 - x )^\sigma ( 1 + k x )$. The best-fit parameters and
the $\chi^2$ per degree of freedom
are quoted in \cite{noi}. In Fig.~\ref{fits}, we present the 
NuTeV data on $F_2(x)$ and $x F_3(x)$ at $Q^2 = 12.59$~GeV$^2$, along 
with the best-fit curves. Similar plots at $Q^2 = 31.62$~GeV$^2$ are 
shown in Ref.~\cite{noi}.

In order to extract individual quark distributions, we need 
to consider also neutral current data. We use BCDMS 
\cite{Benvenuti:1989fm} and NMC \cite{Arneodo:1996qe} results, and 
employ the parametrization of the nonsinglet 
structure function $F_2^{\mathrm{ns}} = F_2^p - F_2^D$ provided by 
Ref.~\cite{DelDebbio:2004qj}. The parametrization~\cite{DelDebbio:2004qj} 
is based on neural networks trained on Monte-Carlo copies of the data set, 
which include all information on errors and correlations: this gives 
an unbiased representation of the probability distribution in
the space of  structure functions.

Writing $F_2$, $x F_3$ and $F_2^\mathrm{ns}$ in terms of their parton 
content, we can extract NLO and NLL-resummed quark
distributions, according to whether we use NLO or NLL coefficient 
functions. We assume isospin symmetry of the sea, i.e. $s = \bar s$ and 
$\bar u = \bar d$, we neglect the charm density, and impose a 
relation $\bar s = \kappa \, \bar u$. We obtain 
a system of three equations, explicitly presented in \cite{noi}, 
that can be solved in terms of $u$, $d$ and $s$. We begin by working
in $N$-space, where the resummation has a simpler form and quark 
distributions are just the ratio of the appropriate structure 
function and coefficient function. We 
then revert to $x$-space using a simple 
parametrization $q(x) = D x^{-\gamma}(1 - x)^\delta$.

\begin{figure}
\centerline{\resizebox{0.48\textwidth}{!}{\includegraphics{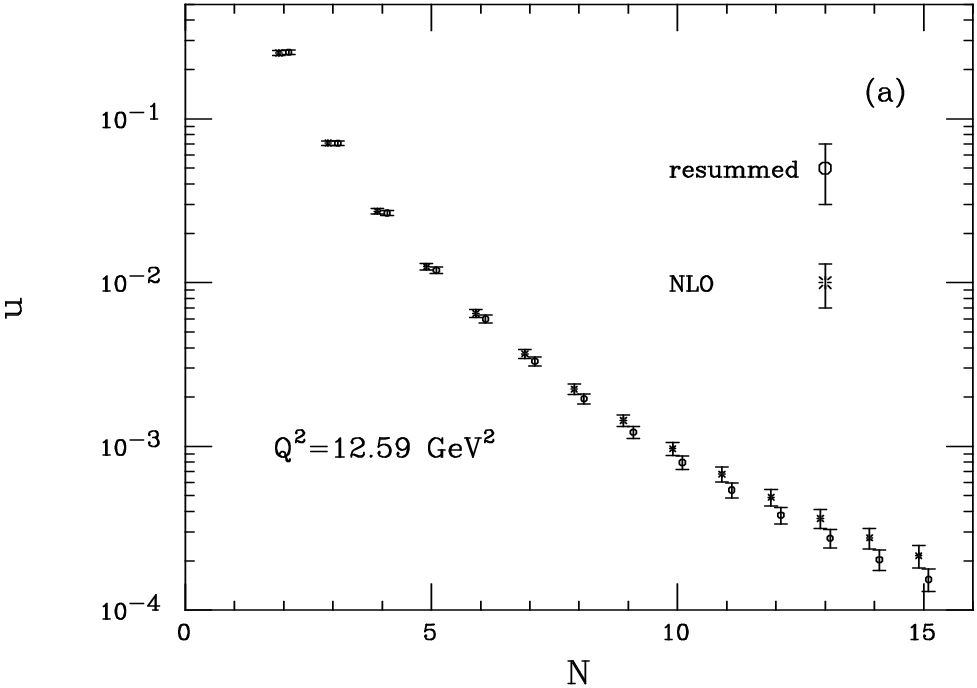}}%
\hfill%
\resizebox{0.48\textwidth}{!}{\includegraphics{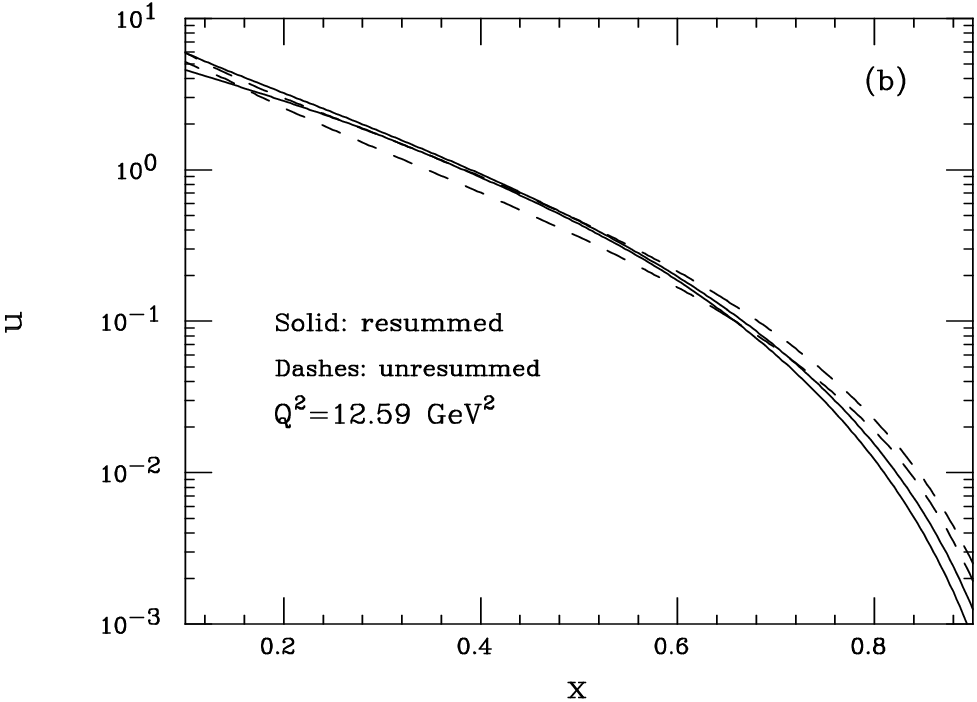}}}
\caption{NLO and resummed up quark distribution at $Q^2 = 12.59$~GeV$^2$
in moment (a) and $x$ (b) spaces. Following \cite{noi}, in $x$ space,
we have plotted the edges of a band corresponding to a prediction at 
one-standard-deviation confidence level (statistical errors only).}
\label{up}
\end{figure}

\begin{figure}
\centerline{\resizebox{0.48\textwidth}{!}{\includegraphics{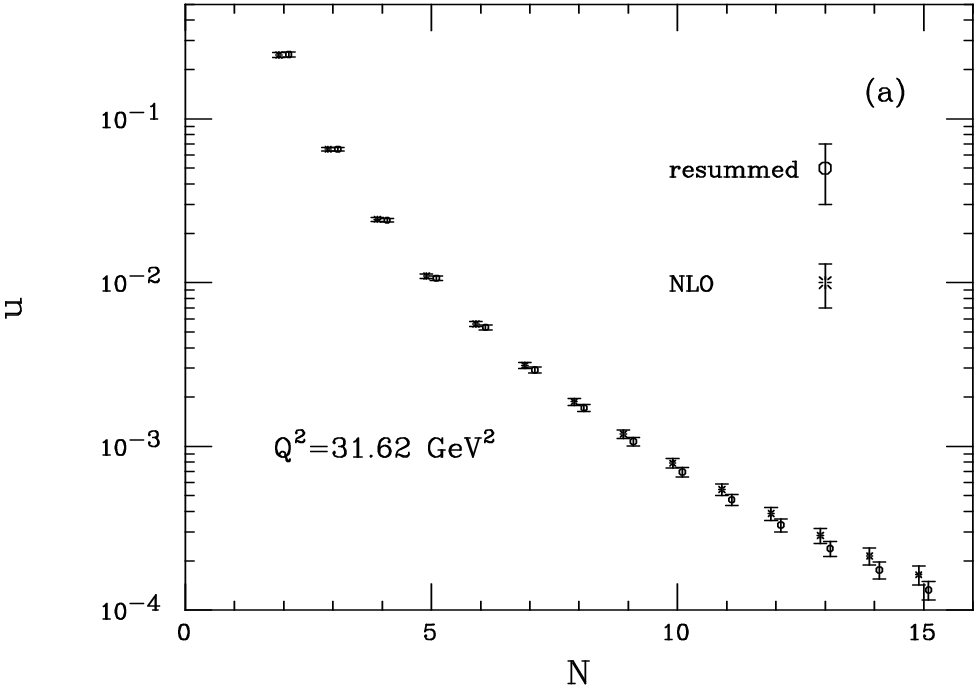}}%
\hfill%
\resizebox{0.48\textwidth}{!}{\includegraphics{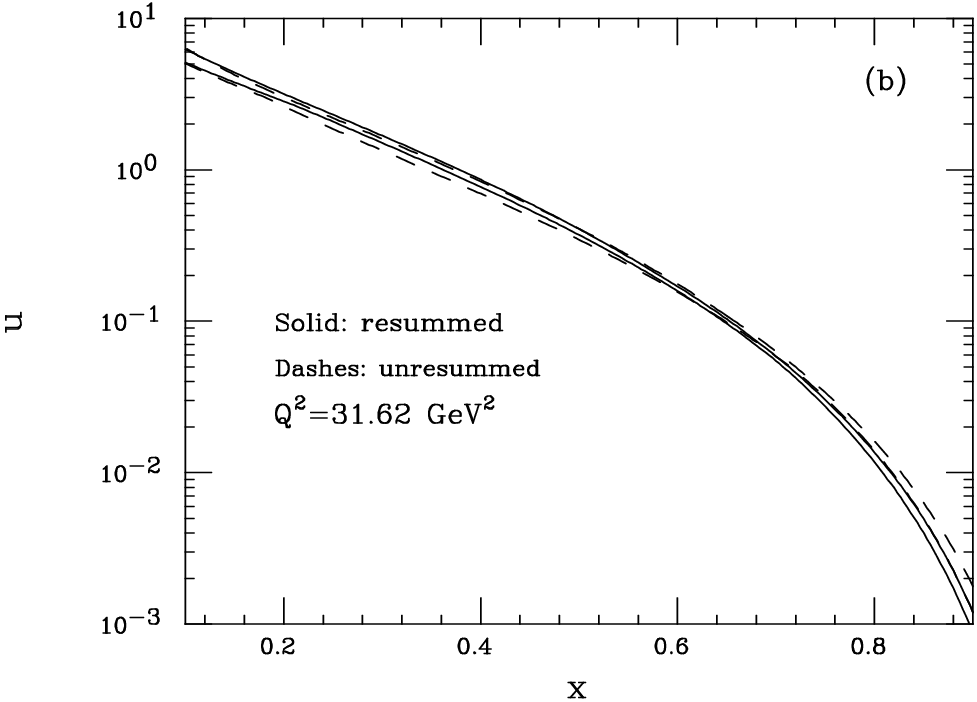}}}
\caption{The same as in Fig.~\ref{up}, but at $Q^2 = 31.62$ GeV$^2$.} 
\label{up1}
\end{figure}

Figs.~\ref{up}--\ref{up1} show the effect of the resummation on the 
up-quark distribution at $Q^2 = 12.59$ and 31.62~GeV$^2$, in $N$- and 
$x$-space respectively. The best-fit values of $D$, $\gamma$ and 
$\delta$, along with the $\chi^2/\mathrm{dof}$, can be found in \cite{noi}.
The impact of the resummation is noticeable at large $N$ and $x$: there,
soft resummation enhances the coefficient function and its moments, 
hence it suppresses the quark densities extracted from structure 
function data. In principle, also $d$ and $s$ densities are affected by 
the resummation; the errors on their moments, however, are too large for
the effect to be statistically significant.
In \cite{noi} it was also shown that the results for the up quark
at 12.59 and 31.62 GeV$^2$ are consistent with NLO perturbative evolution.

In summary, we have presented a comparison of NLO and NLL-resummed quark
densities extracted from large-$x$ DIS data. We found 
a suppression of valence quarks in the $10-20 \%$ range at $x > 0.5$, for 
moderate $Q^2$. 
We believe that it would be interesting and fruitful to extend 
this analysis and include large-$x$ resummation in the toolbox of global 
fits. Our results show in fact that this would be necessary
to achieve precisions better than $10 \%$ in processes involving
large-$x$ partons.



\begin{thebibliography}{99}


\bibitem{Sterman:1986aj} G.~Sterman,
\emph{Nucl.\ Phys.\ B} {\bf 281} (1987) 310.

\bibitem{Catani:1989ne}
  S.~Catani and L.~Trentadue,
  \emph{Nucl.\ Phys.\ B} {\bf 327} (1989) 323.




\bibitem{Laenen:1998kp}
  E.~Laenen and S.~O.~Moch,
  \emph{Phys.\ Rev.\ D} {\bf 59} (1999) 034027.



\bibitem{Corcella:2003ib}
  G.~Corcella and A.~D.~Mitov,
  \emph{Nucl.\ Phys.\ B} {\bf 676} (2004) 346.

\bibitem{Pumplin:2002vw}
  J.~Pumplin, D.~R.~Stump, J.~Huston, H.~L.~Lai, P.~Nadolsky and W.~K.~Tung,
  \emph{JHEP} {\bf 0207} (2002) 012.

\bibitem{Naples:2003ne}
  D.~Naples {\it et al.}  [NuTeV Collaboration],
   hep-ex/0307005.

\bibitem{Yang:2000ju}
  U.~K.~Yang {\it et al.}  [CCFR/NuTeV Collaboration],
  \emph{Phys.\ Rev.\ Lett.\ }  {\bf 86} (2001) 2742.

\bibitem{tzanov}
 M.~Tzanov {\it et al.}  [NuTeV Collaboration], these proceedings.

\bibitem{noi}
 G.~Corcella and L.~Magnea, hep-ph/0506278.

\bibitem{Benvenuti:1989fm}
  A.~C.~Benvenuti {\it et al.}  [BCDMS Collaboration],
  \emph{Phys.\ Lett.\ B} {\bf 237} (1990) 592.

\bibitem{Arneodo:1996qe}
  M.~Arneodo {\it et al.}  [New Muon Collaboration],
  \emph{Nucl.\ Phys.\ B} {\bf 483} (1997) 3.

\bibitem{DelDebbio:2004qj}
  L.~Del Debbio, S.~Forte, J.~I.~Latorre, A.~Piccione and J.~Rojo,
\emph{JHEP} {\bf 0503} (2005) 080.

\end{thebibliography}
\end{document}